\documentclass[11pt,thmsb]{article}%
\usepackage{graphicx}
\usepackage{pdflscape}
\usepackage[english]{babel}
\usepackage{amsmath,amssymb,enumerate,fullpage,setspace}
\usepackage{amsfonts}
\usepackage{color}
\usepackage{geometry}
\usepackage{hyperref}
\usepackage{subcaption, pdflscape}
\usepackage{bigstrut}
\usepackage[authoryear,longnamesfirst]{natbib}
\usepackage{multirow, color}
\usepackage{amsmath}
\usepackage{amssymb}%
\providecommand{\U}[1]{\protect\rule{.1in}{.1in}}
\pdfoutput=1

\newtheorem{assumption}{Assumption}

\newtheorem{condition}{Condition}

\newtheorem{definition}{Definition}

\newtheorem{remark}{Remark}

\DefineNamedColor{named}{green}{RGB}{0,119,75}
\DefineNamedColor{named}{blackred}{rgb}{0.7,0.2,0.4}
\DefineNamedColor{named}{darkblue}{cmyk}{0.8,0.6,0,0.6}
\DefineNamedColor{named}{darkred}{rgb}{0.8,0.1,0.3}
\DefineNamedColor{named}{blackred}{rgb}{0.7,0.2,0.4}
\DefineNamedColor{named}{purplered}{rgb}{0.6,0.2,0.6}
\DefineNamedColor{named}{orange}{cmyk}{0.05,.55,1,0}
\DefineNamedColor{named}{bottlegreen}{rgb}{.3,0.6,.4}
\DefineNamedColor{named}{mediumblue}{RGB}{51,53,178}
\DefineNamedColor{named}{lightblue}{rgb}{.1,.5,.8}
\DefineNamedColor{named}{orangee}{cmyk}{0.04,.05,.12,0}

\renewcommand{\baselinestretch}{1.2}

\numberwithin{theorem}{section}
\begin{document}

\title{\textbf{What Does it Take to Control Global Temperatures?} \\{A toolbox for testing and estimating the impact of economic policies on
climate.}}
\author{\
\begin{tabular}
[c]{ccccc}%
Guillaume Chevillon\thanks{The authors thank participants to the EC$^{2}$
conference in Aarhus, the IAAE at King's College, London, the ISF in Oxford,
the Climate Econometrics online seminar, the SNDE at UCF, Orlando, the
Barcelona Summer Forum, the IWEEE at Bolzano and seminar participants at CREST and Keio University for helpful comments and suggestions. They are in particular
grateful to Jennifer Castle, Frank Diebold, Luca Fanelli, David Hendry, Robert
Kaufmann, Frank Kleibergen, Andrew Martinez, Sophocles Mavroeidis, Zack
Miller, John Muellbauer, Barbara Rossi and Glenn Rudebusch. } &  &  &  &
Takamitsu Kurita\thanks{T. Kurita gratefully acknowledges financial support
from JSPS KAKENHI 24K04877.}\\
{\small ESSEC Business School} &  &  &  & {\small Kyoto Sangyo University}\\
{\small France} &  &  &  & {\small Japan}%
\end{tabular}
}
\date{\today}
\maketitle

\begin{abstract}
{\small \renewcommand{\baselinestretch}{1.05}\selectfont This paper tests the
feasibility and estimates the cost of climate control through economic
policies. It provides a toolbox for a statistical historical assessment of a
Stochastic Integrated Model of Climate and the Economy, and its use in
(possibly counterfactual) policy analysis. Recognizing that stabilization
requires supressing a trend, we use an integrated-cointegrated Vector
Autoregressive Model estimated using a newly compiled dataset ranging between
years A.D. 1000-2008, extending previous results on Control Theory in
nonstationary systems. We test statistically whether, and quantify to what
extent, carbon abatement policies can effectively stabilize or reduce global
temperatures. Our formal test of policy feasibility shows that carbon
abatement can have a significant long run impact and policies can render
temperatures stationary around a chosen long run mean. In a counterfactual
empirical illustration of the possibilities of our modeling strategy, we study
a retrospective policy aiming to keep global temperatures close to their 1900
historical level. Achieving this via carbon abatement may cost about 75\% of
the observed 2008\ level of world GDP, a cost equivalent to reverting to
levels of output historically observed in the mid 1960s. By contrast,
investment in carbon neutral technology could achieve the policy objective and
be self-sustainable as long as it costs less than 50\% of 2008 global GDP and
75\% of consumption.}


\end{abstract}

\newpage

\section{Introduction\label{s: intro}}

This article presents a novel approach to testing the feasibility and
evaluating the costs associated with global temperature control by drawing on
historical data on the interaction between climate and the economy spanning
over a thousand years. Our primary objective is to rely on empirically
\emph{estimated} relationships, acknowledging the criticism raised by Pindyck
(2013\nocite{Pind13}) towards models that solely rely on theoretical
assumptions, calibrations, or simulations, as they may create a
\textquotedblleft misleading perception of knowledge and
precision\textquotedblright. In pursuit of our objective, we construct a
dataset and employ a new econometric methodology relying on stable long-run
interactions to test the hypothesis of temperature controllability and
evaluate its cost through counterfactual policy analysis.

Our approach relies on a model inspired by Stochastic Dynamic Integrated
Models of Climate and the Economy (SDICE) proposed by Nordhaus
(2017\nocite{Nord2017}) that have become one of the main Integrated Assessment
Models (IAMs) of climate and the economy. These models are often assessed
through simulations and scenario analyses that evaluate the cost of carbon
abatement policies required to achieve some specific objectives of temperature
control. The lack of historical carbon abatement policies, at least until the
recent decades, renders empirical studies of the SDICE models limited. This is
one issue we tackle using long historical time series dating back to AD\ 1000,
i.e., before the natural experiment that the industrial revolution
constitutes.\ This allows us to\ prove, through a statistical test, that
policies whose objectives focus on temperature control are empirically feasible.

The industrial revolution having accelerated the upward trend in economic
activity, Greenhouse gas (GHG) emissions and temperatures, we recognize that
temperature control requires suppressing the upward stochastic trend to render
temperatures stable around a long run mean. For this we draw on the abundant
literature studying the dynamic interactions between climate variables and
human activity through an integrated-cointegrated Vector Autoregressive (VAR)
model (see, inter alia, Stern and Kaufmann, 2014\nocite{SterKauf14}, and Chang
\emph{et al.,} 2020\nocite{Chan20}). We study this modeling strategy in light
of work on nonstationary Control Theory developed by Johansen and Juselius
(2001,\nocite{JohaJuse01} 2024\nocite{JohaJuse24}, JJ henceforth) who find
cointegration properties to be the determining factor.\ We show that the
effect of the control policy is to augment the VAR$\left(  p\right)  $ into a
VARMA$\left(  p,1\right)  $, alternatively written as an SVAR$\left(
p\right)  $ with\ recoverable excess shocks that reflect additional, policy
generated, cointegration relations.

Hence, we assess within a cointegrated VAR model what policies are feasible to
render \textquotedblleft climate\textquotedblright\ variables stationary
around a stated objective while retaining economic progress. We focus on very
long series (despite the unavoidable mismeasurements) to capture long-run
equilibria that are invariant to changes in policy. Their stability prior to
and through the industrial revolution constitutes a gauge that these
equilibria are immune to the Lucas (1976\nocite{lucas1976macro}) critique. Our
empirical model can be used as a toolbox for evaluating the \emph{statistical
}feasibility and cost of policies. In an empirical application, we propose a
formal test that carbon abatement or technology investment are capable of
achieving temperature control. We entertain the counterfactual question
whether a centralized authority could have implemented, in the 20th century, a
policy aiming to maintain global temperatures at the level of 1900. We assess
its cost using two distinct \emph{indirect} examples, either $\left(
i\right)  $ via a costing of the lack of carbon abatement policy, using a
reduction of output and consumption as controls; or $\left(  ii\right)  $
through the increased wealth that a costless reduction of the carbon content
of production technology would generate. The estimated cost of the lack of
abatement policy amounts to about 75\% of the 2008 global level of output
(equivalent to forestalling growth since the 1960s) together with a reduction
of 45\% in consumption. Investment in carbon neutral technology would by
contrast achieve its objective and be self-sustainable as long as it costs
less than 50\% of 2008 global GDP and 75\% of consumption. Both policies show
that, under the condition that the stated temperature control is achieved, the
huge magnitude of investment in mitigating technologies that is required and
is more profitable than a degrowth alternative. We state in the title that our
analysis provides a toolbox as our model and methodology can be applied to
other policies and choices of controls that climate scientists and economists
may prefer.

The rest of this study consists of five sections that we have kept
deliberately short for clarity of the exposition; details, further
explanations and empirical results are provided in an online Supplementary
Appendix. Section 2 briefly reviews\ our choice for the SDICE\ model, then
Section 3 the database we consolidate, and its empirical framing in a
cointegrated VAR\ system. Section 4 discusses the control theory of JJ and
develops some of the results needed for a counterfactual analysis. Section 5
then assesses empirically the controllability of temperature through carbon
abatement policies and performs some counterfactual analyses of its cost.
Section 6 concludes. 

\section{A linearized SDICE Model}

We propose an empirical model for the climate-economy nexus and estimate it
over a thousand years. We introduce a log-linearized SDICE\ that has been
sufficiently streamlined so it can shed light on long-run equilibria estimated
later in the paper. There exist a multiplicity of IAM\ models but most of them
can been seen as refinements or extensions of the main equations we consider
here (see, e.g., Barnett \emph{et al.}, 2022\nocite{BBH22}, and H{\"{a}}nsel
\emph{et al.}, 2022,\nocite{CPC22} for analyses of the uncertainty surrounding
the models). By construction, log-linearization removes the nonlinearities
that are inherent in the discussion surrounding possible future \emph{tipping
points} but these can easily be introduced through additional local trends
(see, e.g., Kim \emph{et al.}, 2020, for an application to climate data\nocite{KOEP20}).

We consider here a simplified version of the SDICE\ model of Nordhaus (2017)
as studied, inter alia., in Ikefuji \emph{et al.}
(2020\nocite{ikefuji2020expected}), see Figure \ref{figSDICE} for a
presentation of its general principles. The key feature is that the model
introduces a negative feedback loop from climate to the economy. At each
period $t,$ human economic activity combines various production factors (such
as labor force and capital) to generate real Gross Domestic Product (GDP, or
world output),\ $Y_{t}.$ This production generates externalities in the form
of greenhouse gas (GHG) emissions -- Carbon dioxide (CO$_{2}$) in particular.
Humans may decide to mitigate these externalities via abatement, i.e.
investment that reduces the emission producing (\emph{brown}) content of
economic activity. We simplify the model and only present the log-linearized
version of that in Ikefuji \emph{et al.} (2020\nocite{ikefuji2020expected}),
with lower case letters representing logarithms, see the Supplementary
Appendix for more details.

Total CO$_{2}$ emissions consist of anthropogenic emissions (caused by human
activity) and other -- exogenous -- types $e_{t}^{0}$. Total emissions $e_{t}$
then result from
\begin{equation}
e_{t}=\sigma_{t}-\mu_{t}+y_{t}+e_{t}^{0},\label{eq: e}%
\end{equation}
where $\sigma_{t}$ is the emissions-to-output ratio for CO$_{2}$ and $\mu_{t}$
is the abatement fraction for CO$_{2}.$ Regarding the presence of CO$_{2}$ in
the atmosphere, the SDICE specifies that its concentration $m_{t}$ accumulates
through interactions with shallow and lower oceans. Thus, $m_{t}$ can be seen
as following a Markov state-space process with two hidden layers (latent
variables). The lack of long historical series on the carbon contents of
oceans leads us to using the reduced form model for $m_{t}$ which follows an
autoregressive model with distributed lags of $e_{t},$ here an ARDL(4,4),
$A\left(  L\right)  m_{t}=B\left(  L\right)  e_{t},$where $L$ denotes the lag
operator such that $Lm_{t}=m_{t-1}$ and $A\left(  \cdot\right)  ,$ $B\left(
\cdot\right)  $ are polynomials of degree 4 (we allow for more flexibility in
the empirical study). 

Now atmospheric temperatures, $h_{t}$ (or their anomalies in lieu of their
logarithm) themselves relate dynamically to atmospheric gas concentrations,
ocean temperatures, as well as extraneous radiative forcing, $f_{t}^{0}$. This
can be written  in log form as an equation linking $h_{t+1}-a_{1}%
m_{t+1}-f_{t+1}^{0}$ to the lagged value of that expression together with
$h_{t}$ and $h_{t-1}$. Finally, we consider the impact of climate on economic
growth: the fraction of GDP not spent on abatement is consumed, $c_{t},$ or
invested, $i_{t},$ along the budget constraint: $y_{t}-\omega_{t}-\xi
h_{t}=\tilde{c}_{t},$ where $\tilde{c}_{t}=c_{t}+i_{t},$ and the logarithm of
the cost of abatement, $\omega_{t},$ satisfies $\omega_{t}=\psi_{t}+\theta
\mu_{t}.$with $\theta>1$ so the cost of abatement increases faster than
abatement itself. Parameter $\xi$ represents damage induced by warming: it
closes the feedback loop from climate to the economy shown in Figure
\ref{figSDICE}. Considering $\psi_{t}$ constant over the historical sample
(but not necessarily in the future, if we consider tipping point scenarios),
we identify $\omega_{t}$ to $\theta\mu_{t}$ below.

The model presented above can be expressed in terms of six endogenous
variables $( y_{t}, \allowbreak\tilde{c}_{t}, \allowbreak m_{t}, \allowbreak
h_{t}, \allowbreak\mu_{t}, \allowbreak\sigma_{t}) $ and two exogenous $\left(
e_{t}^{0},f_{t}^{0}\right)  .$ This leads (removing constant terms and
introducing two lag polynomials $D\left(  \cdot\right)  $ and $G\left(
\cdot\right)  $) to the following equations:
\begin{subequations}
\begin{align}
\text{Economy-Climate nexus}  &  \text{:\quad}A\left(  L\right)
m_{t}=B\left(  L\right)  \left(  \sigma_{t}-\mu_{t}+y_{t}+e_{t}^{0}\right)
,\label{Model1}\\
\text{Carbon-Temperature}  &  \text{:\quad}D\left(  L\right)  h_{t}=G\left(
L\right)  \left(  a_{1}m_{t}+f_{t}^{0}\right)  ,\label{Model2}\\
\text{Damage loop}  &  \text{:\quad}y_{t}=\tilde{c}_{t}+\theta\mu_{t}+\xi
h_{t}. \label{Model3}%
\end{align}

The SDICE\ model is typically solved by the Central Planner who sets the
policy variables: level of abatement, $\mu_{t},$ or its cost $\omega_{t},$ and
carbon content of technology $\sigma_{t}$ to achieve a welfare objective in
terms of $\tilde{c}_{t}.$\footnote{By specifying a production function for
$y_{t}$ as a function of labor and capital, as well as a utility function, the
central planner may consider the tradeoff between consumption and leisure. We
abstain from it for simplification.} We consider this objective from an
empirical perspective below through the concept of controllability, where we
ask whether and what policies can generate a stable equilibrium between
economic activity and climate.\begin{figure}[t]
\centering\scalebox{.65}{\includegraphics{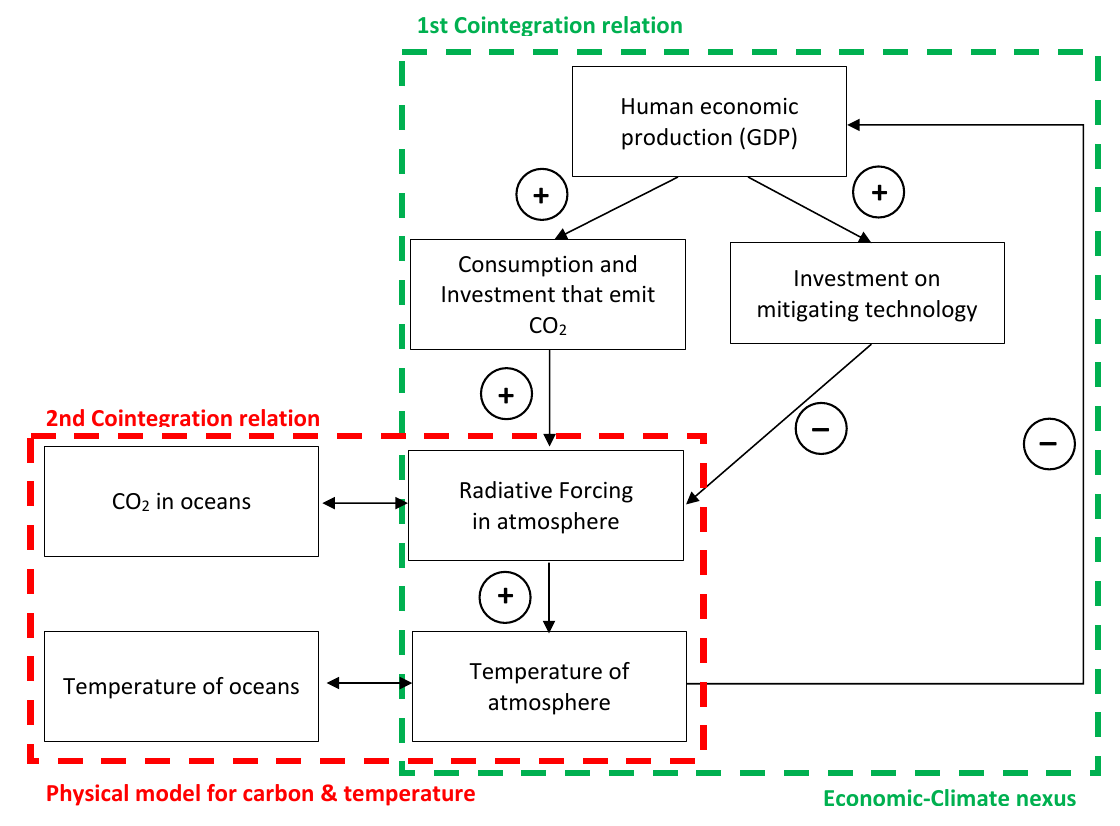}}\caption{Principle
of SDICE Model. The solved out empirical model finds two long term equilibria
( `cointegration relations') corresponding to $\left(  i\right)  $ the
economic-climate nexus and $\left(  ii\right)  $ the physical equilibria
between global temperatures and radiative forcings linked to concentrations of
carbon in the atmosphere and the oceans.}\label{figSDICE}.%
\end{figure}

\section{An Empirical Model for Climate and the Economy}

The dynamic model described above can be assessed using historical empirical
evidence through a cointegrated vector autoregressive model (CVAR) developed
by Johansen (1988\nocite{johansen1988statistical}), i.e., a model for the
dynamic interactions and long run equilibria between climate of and the
economy. For this we first construct a new dataset compiling and extending
various sources over the second millennium AD.

\paragraph{Constructing a long dataset.}

The data were obtained and reconstructed from various sources that are
presented in the Supplementary Appendix, and whose online links and
interpolation details are provided in separate code file (\emph{Data
Preparation}). The data is presented in Figure \ref{figdata}: it comprises
measures of economic activity (real world output and consumption) as well as
variables describing temperature anomalies, carbon concentrations in the
atmosphere, radiative forcings of non-anthropogenic origin (solar,
volcanic...). We consolidate this data at the annual frequency, dating back to
year 1000\ AD.

It is obvious that the data we consider must necessarily be subject to
mismeasurement as we consider global variables over extended periods of time.
Yet, following Duffy and Hendry (2017\nocite{DuffHend17}), we expect that this
mismeasurement does not affect inference on the presence of cointegration. In
Figure \ref{figdata}, we see that all variables except radiative forcings of
volcanic origin exhibit upward trends towards the end of the sample.

\begin{figure}[p]
\centering\scalebox{.7}{\includegraphics{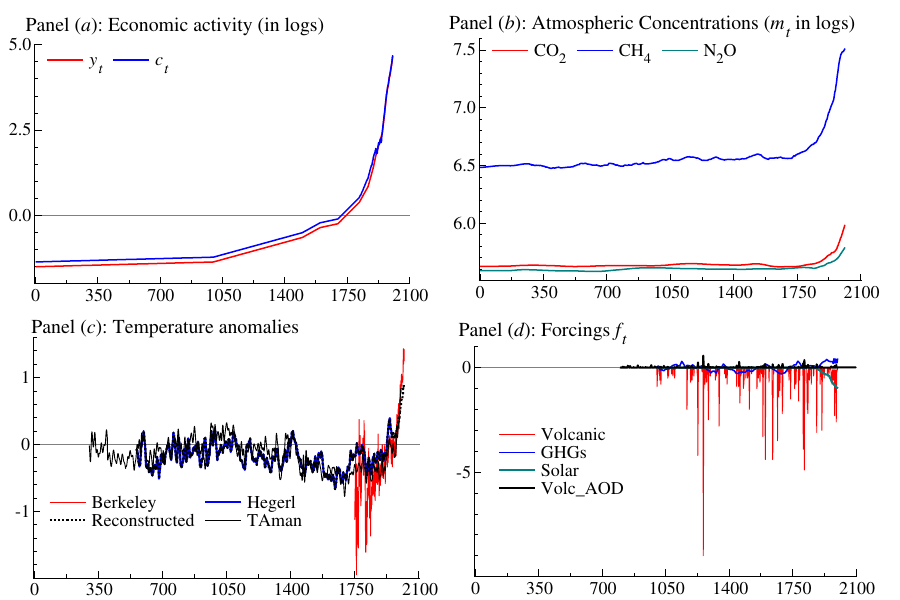}}\caption{Data series
obtained from various sources. All references and explanations are the
Supplementary Appendix.}\label{figdata}%
\end{figure}

\paragraph{The Cointegrated VAR\ approach.}

We perform a cointegration analysis within a VAR$\left(  8\right)  $ model for
the four variables $X_{t}=\left(  y_{t},c_{t},m_{t},h_{t}\right)  $ measured
as log GDP and log consumption, log CO$_{2}$ atmospheric concentration and
actual (non-logged) temperature anomalies. The impact of the Industrial
Revolution is captured by a broken linear trend that starts in\ 1800 (it is
restricted to the cointegrating space). We prefer to model long term
population patterns through this trend for statistical reasons, as
conditioning on the actual data impacts estimator distributions in a
non-standard way. For technical reasons, we also need to include the change in
the broken trend, i.e., a step dummy starting in 1800. Solving the model for
concentrations and radiative forcings as proxies for emissions allows to
extend the data set to cover the years 1000-2008. Our analysis leads to
focusing on $X_{t},$ with radiative forcings of volcanic origin, $f_{t}^{Vol}$
entering as an unrestricted exogenous regressor.

The cointegrated model writes as
\end{subequations}
\begin{equation}
\Delta X_{t}=\tau+\delta_{0}1_{\left\{  t\geq1800\right\}  }+\alpha\left(
\beta^{\prime}X_{t-1}+\delta_{1}t1_{\left\{  t\geq1800\right\}  }\right)
+\sum\limits_{i=1}^{7}\Gamma_{i}\Delta X_{t-i}+\gamma^{\prime}\Delta
f_{t-1}^{Vol}+\epsilon_{t},\label{CVAR}%
\end{equation}
where $\beta^{\prime}X_{t-1}$ captures the two long run equilibria taking the
form of \textquotedblleft cointegration\textquotedblright\ relations that are
stationary although the individual variables themselves are not.

In the empirical model, the data indicates the presence of two stochastic
trends, and two cointegration relations. Based on the SDICE model, we make the
following assumption on the common stochastic trends.

\begin{assumption}
[S]The sources of nonstationarity in the empirical system come from two latent
common stochastic trends:\newline$\left(  i\right)  \quad$a measure of the
\emph{carbon content of technological progress}, which in the model amounts to
$\sigma_{t}$, the logarithm of emissions-to-output ratio of CO$_{2};$
\newline$\left(  ii\right)  \quad$a \emph{wealth effect} that combines global
population with capital accumulation through investment, $i_{t}.$
\end{assumption}

\begin{remark}
We could alternatively interpret a linear combination of them as wealth and
technology of \emph{green} (carbon free) and \emph{brown} (carbon intensive)
origin and impact, noting that these are not orthogonal in the sample.
\end{remark}

\noindent Assumption S helps us identify the two stable (i.e. stationary) long
run cointegration relations (see the Supplementary Appendix for a description
of the estimated model) as
\begin{align*}
c_{1,t}  &  =y_{t}-1.54c_{t}-.81h_{t}-.036t\times1_{\left\{  t\geq
1800\right\}  },\\
c_{2,t}  &  =h_{t}-4.12m_{t}.
\end{align*}
We interpret them with the help of the SDICE model.

\begin{enumerate}
\item The first equation, $c_{1,t}$, corresponds to an interaction between
temperature and human activity. In the SDICE, this corresponds to equation
(\ref{Model3}).

\item The second cointegrating relation, $c_{2,t}$, corresponds to the
interplay between CO$_{2}$ concentrations in the atmosphere and temperature,
as in\ equation (\ref{Model2}).

\item The data does not support equation (\ref{Model1}) as a cointegration
relation. This is in line with Assumption S.$\left(  i\right)  $ that includes
the nonstationary $\sigma_{t}.$
\end{enumerate}

The second cointegration relation does not involves human activity although
the technology mix has strong evolved over the millennium. Yet, $y_{t}$ is
close to significant in this equation so we might have preferred to retain it.
In Smil (2017, in his last table on page 458\nocite{Smil2017}) who reports
estimates of per capita annual consumption of primary energy, these have only
started increasing by an order of magnitude when countries started their
industrial revolution (the estimates are explicitly uncertain). Hence, it is
likely that the technology mix, prior to 1800, has not modified much the
elasticity of temperature to CO2 emissions, yet given the ensuing changes, we
prefer to leave human activity outside this equation, to reflect only physical
equilibria. For robustness, we reestimated the model starting in AD\ 1750, and
found the estimates to be very similar, with a test of overidentifying
restrictions that has a $p$-value of 0.41.

In the model above, 100\%\ green growth is feasible if a downward trend in
$\sigma_{t}$ is achieved, tending towards $-\infty$ so actual carbon content
$\Sigma_{t}=\exp\left(  \sigma_{t}\right)  $ is driven to zero. This
constitutes an alternative to reducing growth altogether. In practice, given
the current technologies, a combination of \emph{green} investment and
restrained \emph{brown} growth is required in our model to achieve stability
in temperature, see the counterfactual analysis we perform below.

\paragraph{Comparison with estimates in the literature.}

To assess the plausibility of our estimated model, we compare its implications
with meta analyses based on the existing literature. The long run impact of
climate change on the economy has been considered nonlinear so calibrated
parameter uncertainty is an important issue.\ The table below summarizes the
results for three main indicators used in the climate-economy literature (see
the Appendix for derivations and explanations).

\begin{center}%
\begin{tabular}
[c]{lcc}\hline\hline
& \text{Ranges in Literature} & \text{Our results }$\left(  s.e.\right)
$\\\hline
\text{Temperature damage on GDP, }$\xi$ & $\left[  .68,1.34\right]  ^{\ast}$ &
$.81$\text{ }$\left(  .40\right)  $\\
$200$\ \text{year temperature increase due to CO}$_{2}$ & $\left[
.7,2.1\right]  ^{\dag}$ & $1.35$\text{ }$\left(  .66\right)  $\\
\text{GDP loss due to CO}$_{2}$\text{, }$\gamma$ & $\left[  .27,10.4\right]
^{\dag}$ & $4.25$\text{ }$\left(  .38\right)  $\\\hline\hline
\end{tabular}
\newline$^{\ast}${\small Nordhaus \& Moffat (2017\nocite{NorMof17}); }$^{\dag
}${\small Hassler, Krusell \& Olovsson (2018\nocite{HKO18}).}
\end{center}

Overall, these results (with reported standard errors) show that our estimates
are very much in the low to mid-range of those reported in the literature, so
we are confident that our analysis neither severely underestimates nor
overestimates the relative impacts of climate and the economy.

\section{Control theory in a cointegrated system\label{s: theo}}

We now review the non-stationary control theory derived by JJ and show how it
can be used to understand the issue of climate control through carbon
abatement, reinterpreting the objective of the policy as suppressing the
stochastic trend in temperature through use of cointegration properties. For
the sake of expositional simplicity, setting $k=1$ in (\ref{CVAR}) and
removing deterministic terms reduces the model to
\begin{equation}
\Delta X_{t}=\alpha\left(  \beta^{\prime}X_{t-1}-\mu\right)  +\epsilon
_{t},\label{VEC}%
\end{equation}
and its Granger-Johansen moving average representation is
\begin{equation}
X_{t}=C%
{\textstyle\sum\nolimits_{j=1}^{t}}
\epsilon_{j}+C\left(  L\right)  \epsilon_{t}+A_{0},\label{VMA}%
\end{equation}
where the long-run impact matrix $C$ governs the nonstationary stochastic
trends driving the nonstationary system: it plays a key role below in the
theory of controllability. In the expression above, $C\left(  L\right)
\epsilon_{t}$ represents a stationary series, $A_{0}\ $depends upon the
initial values and $\mu$ such that $\beta^{\prime}A_{0}=\mu$. It follows from
$\beta^{\prime}C=0$ that $\beta^{\prime}X_{t}-\mu=\beta^{\prime}C\left(
L\right)  \epsilon_{t}$ is stationary so $\beta$ are cointegration vectors.
The long-run expected value of $X_{t}$ is defined as $X_{\infty}=\lim
_{\tau\rightarrow\infty}\mathsf{E}\left(  X_{\tau}\left\vert X_{0}\right.
\right)  =CX_{0}+\alpha\left(  \beta^{\prime}\alpha\right)  ^{-1}\mu.$ We now
consider a Control Theory derived from Preston and Pagan (1982, Chapter 4, and
Section 5.8 in particular\nocite{pagan1982theory}) and adapted to the
nonstationary context as follows.

\begin{definition}
The \emph{Control Policy} consists in two selection matrices $\left(
a,b\right)  $, an objective $b^{\ast}$ and a contemporaneous control rule
$\nu\left(  \cdot\right)  $ such that, in the system (\ref{VEC}),
\newline\noindent$\left(  i\right)  $ \emph{Policy controls} $a^{\prime}X_{t}$
can be changed by intervention using control rule $\nu\left(  X_{t}\right)  :$%
\[
\mathsf{ctr}:a^{\prime}X_{t}^{ctr}=a^{\prime}X_{t}+a^{\prime}\nu\left(
X_{t}\right)  ;
\]
\noindent$\left(  ii\right)  $ The \emph{objective} is the desired value
$b^{\ast}$ of a\ \emph{targeted} combination of variables, $b^{\prime}X_{t}.$
It is defined as the long-run conditional expectation%
\[
b^{\ast}=\lim_{h\rightarrow\infty}\mathsf{E}\left(  b^{\prime}X_{t+h}%
^{new}|X_{t}^{ctr}\right)  ,
\]
where $b^{\prime}X_{t+1}^{new}$ is the ecosystem outcome for $X_{t+1}$ (now
made stationary) given by expression (\ref{VEC}) when the control has been
applied to $X_{t}.$
\end{definition}

We assume in this paper that the policy objective is to control temperature so
that $b^{\prime}X_{t+1}^{new}=h_{t+1}^{new}$ and this process becomes
stationary around a mean $b^{\ast}$.

The control policy defined above is explicitly written as a new equation to
the system, one which requires that the \textquotedblleft
authority\textquotedblright\ that implements it must be able to modify
$a^{\prime}X_{t}$ via $a^{\prime}\nu\left(  X_{t}\right)  .$ This may require
extra controls that are outside the system but interact with it, though for
simplicity here we disregard this possibility. Indeed, our aim is not to
assess \emph{how }to implement a policy, but \emph{whether} it can be effective.

With continuous monitoring and control, the procedure delineated above works
as follows (for a policy that starts at time $t=0$)%
\[
\color{black}X_{0}\color{blue}\rightarrow\underset{\color{blue}(\text{Policy}%
)}{\underbrace{X_{0}^{ctr}=X_{0}+\nu\left(  X_{0}\right)  }}%
\color{red}\rightarrow\underset{\color{red}(\text{Ecosystem})}{\underbrace
{X_{1}^{new}=\left(  I_{p}+\alpha\beta^{\prime}\right)  X_{0}^{ctr}%
-\alpha^{\prime}\mu+\varepsilon_{1}}}\color{blue}\rightarrow\underset
{\color{blue}(\text{Policy})}{\underbrace{X_{1}^{ctr}=X_{1}^{new}+\nu\left(
X_{1}\right)  }}\color{red}\rightarrow....
\]

\paragraph{Invariance.}

Success of the above approach relies on the notion of \emph{invariance} of the
system (\ref{VEC}) to interventions of the type $a^{\prime}X_{t}\rightarrow
a^{\prime}X_{t}^{ctr}$ so that we can assume that the potential outcome is
generated through mechanism (\ref{VEC}) that is not affected by
interventions.\ For invariance, we hence require here that parameters remain
unaffected by the introduction of the new policy so we can be assured that
$X_{t+1}^{new}=X_{t}^{ctr}+\alpha\left(  \beta^{\prime}X_{t}^{ctr}-\mu\right)
+\epsilon_{t+1}$.

This relates to the notion of \textquotedblleft
super-exogeneity\textquotedblright\ proposed by Engle, Hendry and Richard
(1983\nocite{engle1983exogeneity}), and studied by Pretis
(2021\nocite{PRETIS21}) in the context of climate. Yet, super-exogeneity is
about invariance of conditional equations, while we consider here a system
approach. While invariance cannot be ascertained with certainty, we follow in
the Supplementary Appendix two approaches to ensure it constitutes a plausible
assumption. These rely on $(i)$ estimating the model using extra long
historical data to capture stable relations pre- and post-industrial
revolution, treating the latter as an historical \textquotedblleft natural
expirement\textquotedblright\ in climate change; and $(ii)$ using statistical
tests used in the context of super-exogeneity, inter alia by Castle et al.,
(2017\nocite{CHM17}).\ 

\paragraph{Policy Impact.}

A question\ raised by JJ is that of controllability of $b^{\prime}X_{t}$ via
$a^{\prime}X_{t}.$ Since the objective is formulated in terms of a conditional
expectation for $b^{\prime}X_{t+h}^{new}$, the choice of controls and policy
rule must ensure that $b^{\prime}X_{t+h}^{new}$ is indeed stationary around
$b^{\ast}.$ In the cointegrated VAR(1), JJ show that
\[
b^{\ast}=b^{\prime}\underset{h\rightarrow\infty}{\lim}\mathsf{E}\left(
X_{t+h}^{new}\left\vert X_{t}^{ctr}\right.  \right)  =b^{\prime}\left(
C\left[  X_{t}+\nu\left(  X_{t}\right)  \right]  +\alpha\left(  \beta^{\prime
}\alpha\right)  ^{-1}\mu\right)  ,
\]
so that the condition for controllability writes as follows.

\begin{condition}
[Controllability, \textsf{C}]The Policy objective is achievable using the
chosen controls if
\begin{equation}
\det\left(  b^{\prime}Ca\right)  \neq0. \label{eq: condition C}%
\end{equation}
where $C$ is the matrix measuring the long run impact of the stochastic trends
in the Granger-Johansen moving average representation (\ref{VMA}).
\end{condition}

Notice that Controllability is a property of the system (under invariance) for
the targeted variables and chosen controls. It does not depend on the actual
rule $\nu\left(  X_{t}\right)  .$ JJ show that if Controllability applies,
then a linear rule achieves it:%
\begin{equation}
\nu\left(  X_{t}\right)  =a\left(  b^{\prime}Ca\right)  ^{-1}\left[
\underset{\text{policy discrepancy}}{\underbrace{\left(  b^{\ast}-b^{\prime
}X_{t}\right)  }}+b^{\prime}\alpha\left(  \beta^{\prime}\alpha\right)
^{-1}\underset{\text{system discrepancy}}{\underbrace{\left(  \beta^{\prime
}X_{t}-\mu\right)  }}\right]  \equiv\overline{a}\left(  \kappa^{\prime}%
X_{t}^{new}-\kappa_{0}\right)  ,\label{nu rule}%
\end{equation}
where we define the projector onto the space spanned by $a$ as $\overline
{a}=a\left(  a^{\prime}a\right)  ^{-1}.$ The rule consists of a weighted
average of $b^{\ast}-b^{\prime}X_{t}$, a \textquotedblleft
policy\textquotedblright\ discrepancy between the desired objective and the
current value at $t,$ and $\beta^{\prime}X_{t}-\mu$ is a \textquotedblleft
system\textquotedblright\ deviation from the steady state at $t.\ $Policy
becomes, here, fully endogenous and does not constitute an exogenous shock, as
is often modelled in economics via structural VARs (more on this below). The
reason for the effectiveness of the policy, and hence the channel through
which it operates, relies on $\nu\left(  X_{t}\right)  $ being stationary and
generating an extra linear cointegration relation. The\emph{ new} augmented
system writes,%

\begin{equation}
\Delta X_{t+1}^{new}=\alpha\left(  \beta^{\prime}X_{t}^{new}-\mu\right)
+\left(  I_{p}+\alpha\beta^{\prime}\right)  \nu\left(  X_{t}^{new}\right)
+\varepsilon_{t+1}=\left(  \alpha,\left(  I_{p}+\alpha\beta^{\prime}\right)
\overline{a}\right)  \left[
\begin{array}
[c]{c}%
\beta^{\prime}X_{t}^{new}-\mu\\
\kappa^{\prime}X_{t}^{new}-\kappa_{0}%
\end{array}
\right]  +\varepsilon_{t+1} \label{VECnew}%
\end{equation}

In the context of a VAR$\left(  1\right)  $ dynamic system, it can be shown
that $\nu\left(  X_{t}^{new}\right)  =\overline{a}\kappa^{\prime}%
\varepsilon_{t},$ so the impact of the policy is to augment the
\color{blue}VAR$\left(  1\right)  $ \color{black}into a
\color{red}VARMA$\left(  1,1\right)  \color{black}:$%
\begin{equation}
\color{blue}X_{t+1}^{new}=-\alpha\mu+\left(  I+\alpha\beta^{\prime}\right)
X_{t}^{new}+\varepsilon_{t+1}+\color{red}\left(  I+\alpha\beta^{\prime
}\right)  \overline{a}\kappa^{\prime}\varepsilon_{t}%
\color{black}.\label{SVARMA}%
\end{equation}
\color{black} This results also holds for higher order VAR$\left(  p\right)  $
dynamics that are modified into VARMA$\left(  p,1\right)  $ (for a careful
choice of policy parameters among those that achieve the stated objective, see
JJ, Theorem 6).\ Equation (\ref{SVARMA}) shows that the policy can be
identified in practice for its parameters $\overline{a}\kappa^{\prime}$
through a Structural VARMA or VMA and associated response function. Following
the most recent literature on treatments in time series and macroeconometrics,
the policy controls and objective above define a Direct Potential Outcome
System in the sense of Rambachan and Shephard (2021\nocite{rambashep21}). In
their framework, the policy consists in an assignment made at time $t+1$,
$W_{t+1}=\nu\left(  X_{t}^{new}\right)  $ that is uncorrelated with
$\varepsilon_{t+1}$ so
\[
X_{t+1}^{new}=-\alpha\mu+\left(  I+\alpha\beta^{\prime}\right)  X_{t}%
^{new}+\left(  I+\alpha\beta^{\prime}\right)  W_{t+1}+\varepsilon_{t+1}%
\]
constitutes a \textquotedblleft\emph{Non-anticipating Potential Outcome}%
\textquotedblright\ (see the Supplementary Appendix for a discussion of their
framework and ours). Alternatively, an SVAR representation exists where the
assignment $W_{t+1}$ constitutes an \textquotedblleft excess
shock\textquotedblright\ (see Pagan and Robinson,
2022\nocite{PaganRobinson2022}) that is chosen, orthogonally to $\varepsilon
_{t+1}$, to coincide with $\overline{a}\kappa^{\prime}\varepsilon_{t}.$ Since
$\overline{a}\kappa^{\prime}$ is chosen to add a stationary relation in the
system, the MA term in equation (\ref{SVARMA}) is invertible and $W_{t+1}$
recoverable from past observations (Chahrour and Jurado,
2022\nocite{ChahJura22}). Such an analysis does not consitute our objective
here though, since our aim is to study the impact of implementing a given
policy. Yet, we retain the Rambachan and Shephard interpretation that the
process observed after the introduction of the sustained control policy is
$X_{t}^{new},$ not $X_{t}^{ctr}$ as may seem to result from the definition above.

\paragraph{Testing for Climate Controllability.}

Our first question is whether economic activity, say $y_{t}$, $c_{t}$ or
$m_{t},$ can be used as an instrument for a policy that aims to control
temperature. Based on the analysis above which can be readily extended to
cover $k>1$, this amounts to testing the significance of the element in the
long run $C$ matrix that enters Condition \textsf{C}. 

The matrix estimate and corresponding $t$-statistics show that the null of non
controllability of temperature $h_{t}$ -- i.e. $b^{\prime}Ca=0$ in equation
(\ref{eq: condition C}) -- by each of the potential policy controls $y_{t}$,
$c_{t}$ or $m_{t}$ (or a linear combination thereof) can safely be rejected at
conventional levels. This shows that carbon abatement can achieve its purpose
of controlling temperature, i.e. rendering it stationary around a chosen mean.
Such policies can amount to carbon mitigation through economic adaptation
(controls $y_{t}$ and $c_{t}$), or via direct capture of carbon dioxide in the
atmosphere (control $m_{t}$), provided the technology develops sufficiently
fast. Alternative choices of policies can be introduced through the various
equations of the model (see, e.g., Policy\ 2\ below). 

\section{Counterfactual and Prospective Policy Analysis}

Now that we have established that a policy of carbon abatement is capable of
controlling global temperatures, the natural follow-up question is at what
cost. To this end, this section considers the retrospective and prospective
costs of a policy. For this, we design, and then simulate using the empirical
model, a counterfactual path for the endogenous variables.

\paragraph{Policy design.}

We now consider JJ's analysis from the perspective of the policy maker who
aims to perform a historical counterfactual analysis. We assume that the
policy is actioned by a central authority, which may represent international
coordination, or might be fictitious. For instance, a counterfactual analysis
that may be of interest consists in deriving the development path that would
have arisen if mitigating policies had been put in place through carbon
abatement at some point in the past.

The methodology and model above allow for a variety of policy objectives and
instruments. Here, we consider retrospective policies that would have aimed to
control global temperatures and to render them stationary around a long run
mean equal to their 1900 level (assuming $h_{t}$ were observed then), i.e.
stable over the 20th century at 0.7$%
{{}^\circ}%
$C below their 2008 level. Naturally, the ensuing cost depends on the timing
of the policy initiation as well as on the choice of controls.

We contemplate two distinct policies that provide different approaches to
measuring the opportunity cost of inaction on green investment.

\begin{enumerate}
\item One that mixes investment and consumption, with more weight accrued to
the latter (since our model does not specify the increase in abatement
effectiveness that would have arisen from higher investment on mitigating
technologies -- yet this can be easily embedded via specific assumptions). Our
baseline choice for the control is $a^{\prime}X_{t}=y_{t}+c_{t}/2$ with target
$b^{\prime}X_{t}=h_{t}$. The weights are arbitrary but the modeler has the
flexibility to consider policies of their choice, as our study provides a
\emph{toolbox} that is adaptable.

\item Another that targets the same objective through a reduction of the
carbon content of technology, $\sigma_{t},$ in equation (\ref{Model1}),
assuming that it can be put in place to control emissions and concentrations
$a^{\prime}X_{t}=m_{t}$ directly. This amounts, for a given level of economic
activity, to reducing emissions and, possibly, to capturing atmospheric
CO$_{2}.$ Notwithstanding the actual cost of developing these technological
solutions -- which is assumed to be zero in a baseline scenario, but could be
introduced as a fraction of $y_{t}$ or $\tilde{c}_{t}$ -- such a policy would
restrain the economic damage caused by increasing temperatures, hence
increasing $y_{t}$ and $\tilde{c}_{t}.$ As we do not model the cost of such a
reduction, the ensuing impact on $y_{t}$ and $c_{t}$ can help the policy maker
assess what costs they are willing to endure for a specific objective.
\end{enumerate}

\paragraph{Policy 1: Cost of inaction on abatement technologies.}

\begin{figure}[ptb]
\centering$%
\begin{array}
[c]{cc}%
\hspace{-.7cm}\text{\scalebox{.55}{\includegraphics{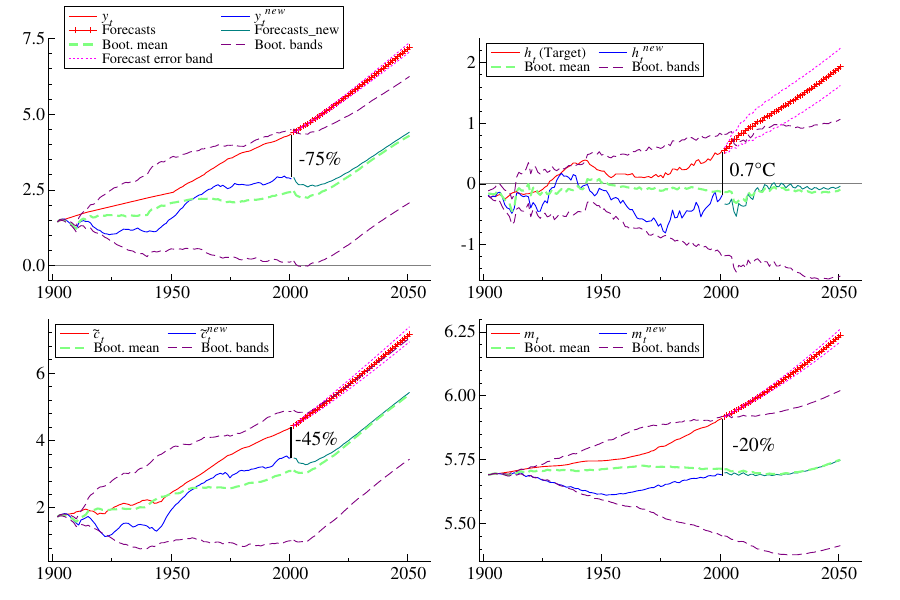}}}
& \scalebox{.55}{\includegraphics{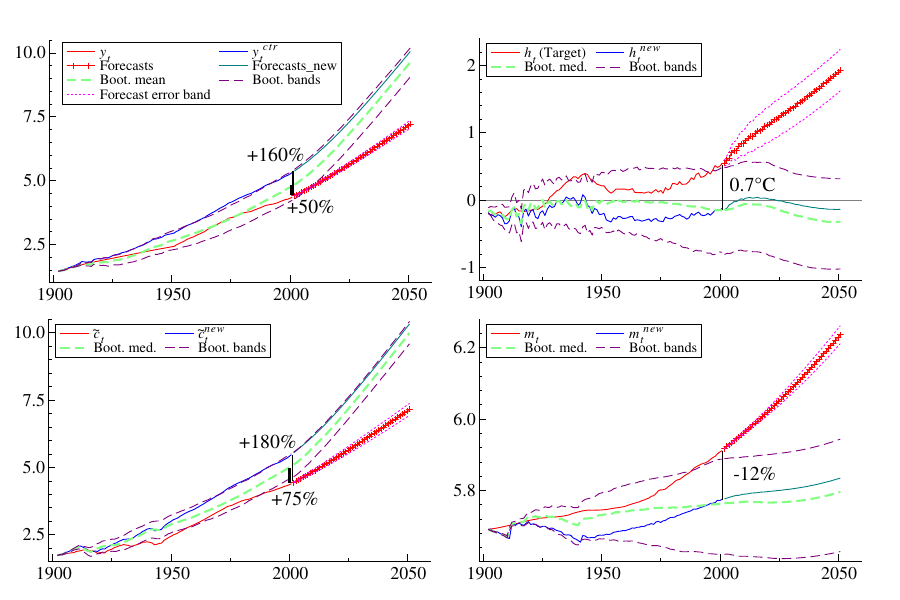}}\\
\text{Panel }\left(  a\right)  & \text{Panel }\left(  b\right)
\end{array}
$ \caption{Counterfactual cost of a policy starting in 1900\ aiming at
controling global temperatures and rendering them stationary around a long run
mean equal to their 1900 level (as it is known now). Panel $\left(  a\right)
$: the control is $y_{t}+\tilde{c}_{t}/2.\ $Panel $\left(  b\right)  $: the
control is $m_{t}.$}%
\label{fig: Policy1900}%
\end{figure}

Assuming an authority has modified at will both world GDP and consumption to
achieve its objective, Figure \ref{fig: Policy1900}, Panel $\left(  a\right)
$, reports the resulting dynamics, where to avoid a sudden shock in the early
20th century, we introduce the policy progressively. We also include a
forecast over the first half of the 21st century, both as obtained
unconditionally from the empirical model and with a policy that aims to
maintain the same temperature level. We also produce the bootstrap mean
prediction and associated confidence intervals, restricting ourselves to using
residuals post 1900.

Temperature control is achieved in this exercise via stabilizing\ atmospheric
carbon concentrations to a level about 20\% below that of 2008. The ensuing
cost in terms of foregone GDP\ in 2008 is about 75\% so the observed
counterfactual GDP in\ 2008 would have been that which we have known in the
1960s (the uncertainty is large), i.e., a cost of about 40 years of growth.
The cost in terms of world consumption is 45\% of the 2008 level, foregoing
the growth observed since the mid-1970s. When looking at the bootstrap
distributions (at each step forecasting the next period using 500 bootstrap
samples), we see that the historical sequence of shocks imposed that most of
the gains, in the counterfactual experiments, where obtained in the second
half of the 20th century.

In order to assess the cost of inaction in the face of climate change, we also
perform a complementary analysis where the objective in terms of temperature
control remains the same, but the policy only starts in 1950. Corresponding
counterfactual outcomes are presented in the Supplementary Appendix. The
ensuing cost becomes 90\% of the GDP of 2008, i.e. essentially no growth since
1950. In terms of consumption, the counterfactual stands at 75\%\ below the
observe level of 2008, i.e. an additional 20\%\ reduction to the baseline scenario.

Projecting our experiment over the 21st century in either policy, we see that
the efforts will have to be sustained, reinforcing abatement policies.
Clearly, these projections are contingent on specific assumptions over the
forecast period and nonlinearities due to major climate change (see, e.g.,
Diebold \emph{et al.}, 2023,\nocite{Dieb23} and Lenton \emph{et al.},
2019\nocite{lenton2019climate}).

\paragraph{Policy 2: Reducing the carbon content of technology}

Assuming the technological improvement is available freely but activated
progressively to reach a reduction in 20\% of atmospheric CO$_{2}$
concentrations, the resulting gain in economic activity is potentially
massive: Figure \ref{fig: Policy1900}, Panel $\left(  b\right)  ,$ shows that
the bootstrap interval ranges from $-3\%\ $to +$160\%$, with a mean gain of
+50\%, and where the realized policy is at the upper bound. Similar values
hold for $\tilde{c}_{t},$ with a wider bootstrap range and higher mean
(+75\%). These positive gains show that there exists a rationale for massive
\emph{self-financing} investment in \textquotedblleft green\textquotedblright%
\ technologies.\ Uncertainty is also high here, yet much less so than via
Policy 1. 

\section{Conclusions}

This paper assesses the feasibility and quantifies the cost of carbon
abatement policies using long series of economic and climate data compiled for
the second millennium AD. By means of a cointegrated VAR modeling strategy
that matches a simple linearized SDICE\ model, we test whether and show how a
policy that aims to render temperatures stationary around a given long run
mean can be achieved. In an empirical application, we test that carbon
abatement is indeed significantly capable of such a policy. We assess its
counterfactual cost, if a centralized authority had been able to implement
such a policy over the 20th century to maintain the globe's temperature at its
level of 1900 (observed ex post). These costs are first assessed under the
assumption of a constant carbon content of technology, so they show the
opportunity cost of the lack of investment in mitigating technologies. This is
corroborated by a complementary counterfactual policy where carbon emissions
and concentrations are directly controlable through technology, showing the
massive extent to which technological investment is self-sustainable.

The analysis in this paper constitutes an exercise where we deliberately chose
simple policy controls, but economically meaningful alternatives are also
possible (say the discounted wealth rather than spot GDP and consumption) at
the cost of additional assumptions. Indeed we see the collected data, model
and results above as a toolbox for policy analysis where refinements on
possible projected scenarios and abatement policies need to be assessed. This
constitutes one step further into statistical analyses of the feasibility of
temperature control, and cost assessments of such policies.

\newpage

\renewcommand{\baselinestretch}{1.0} { \selectfont
\bibliographystyle{chicago}
\bibliography{Climatebiblio}
}

\end{document}


\title{Supplementary Appendix to \\``What Does it Take to Control Global Temperatures? A toolbox for testing and estimating the impact of economic policies on climate."}
\author{\
\begin{tabular}
[c]{ccccc}%
Guillaume Chevillon &  &  &  & Takamitsu Kurita\\
{\small ESSEC Business School} &  &  &  & {\small Kyoto Sangyo University}\\
{\small France} &  &  &  & {\small Japan}%
\end{tabular}
}
\date{\today \vspace{-.5cm}}
\maketitle
\tableofcontents

\newpage


\renewcommand{\thesection}{\Alph{section}}

\section{Introduction}

$\allowbreak$This appendix contains descriptions of the model and data,
additional empirical results and further discussions. Equations in this
document are numbered with the prefix `S--'. Equations without prefix refer to
the main paper. This paper uses \textit{CATS }(Doornik and Juselius,
2018\nocite{doornik2018cats}) and \textit{PcGive} (Doornik and Hendry,
2018\nocite{doornik2018oxmetrics}) for the empirical analysis and \textit{Ox}
(Doornik, 2009\nocite{doornik2009object}) for the policy simulations. Here are
some notational conventions we use: let $\rho$ a matrix with full column rank;
its orthogonal complement is denoted by $\rho_{\perp}$ satisfying $\rho
_{\perp}^{\prime}\rho=0$ with the matrix ($\rho,\rho_{\perp}$) being of full
rank. We also define the projector $\overline{\rho}=\left(  \rho^{\prime}%
\rho\right)  ^{-1}\rho^{\prime}.$ $\mathsf{I}(d)$ represent a stochastic
process integrated of order $d$, thus $\mathsf{I}(0)$ indicates a weakly
stationary process.

The sections in this Appendix are referred to in the main text. For clarity of
exposition, we ensure the elements are here sufficiently self-contained to
avoid referring too much to the main text.

\section{Log-linearizing the SDICE\ Model}

Following Ikefuji \emph{et al.} (2020\nocite{ikefuji2020expected}), total
CO$_{2}$ emissions consist of anthropogenic emissions (caused by human
activity) and other -- exogenous -- types $E_{t}^{0}$. Total emissions $E_{t}$
then result from
\begin{equation}
E_{t}=\Sigma_{t}\left(  1-\mu_{t}\right)  Y_{t}+E_{t}^{0}, \label{eq: E}%
\end{equation}
where $\Sigma_{t}$ is the emissions-to-output ratio for CO$_{2}$ and $\mu_{t}$
is the abatement fraction for CO$_{2}.$ Denoting logarithmic transforms by
lower-case letters, the previous equation may be log-linearized as
\begin{equation}
e_{t}=\sigma_{t}-\mu_{t}+y_{t}+e_{t}^{0}, \label{eq: e}%
\end{equation}
where $e_{t}=\log E_{t},$ $\sigma_{t}=\log\Sigma_{t},$ $y_{t}=\log Y_{t}$ and
$e_{t}^{0}=\log\left(  E_{t}/\left(  E_{t}-E_{t}^{0}\right)  \right)  \approx
E_{t}^{0}/E_{t}$ when $E_{t}^{0}<<E_{t}.$

Regarding the presence of CO$_{2}$ in the atmosphere, the SDICE specifies that
its concentration $M_{t}$ accumulates through interactions with shallow and
lower oceans, with concentrations respectively denoted by $M_{t}^{\left(
s\right)  }$ and $M_{t}^{\left(  \ell\right)  }$. For instance, Ikefuji
\emph{et al.} (2020) specify%
\begin{align*}
M_{t+1}  &  =\left(  1-b_{0}\right)  M_{t}+b_{1}M_{t}^{\left(  s\right)
}+E_{t},\\
M_{t+1}^{\left(  s\right)  }  &  =b_{0}M_{t}+\left(  1-b_{1}-b_{3}\right)
M_{t}^{\left(  s\right)  }+b_{2}M_{t}^{\left(  \ell\right)  },\\
M_{t+1}^{\left(  \ell\right)  }  &  =b_{3}M_{t}^{\left(  s\right)  }+\left(
1-b_{2}\right)  M_{t}^{\left(  \ell\right)  .}%
\end{align*}
Given the difficulty in obtaining long historical series on the carbon content
of shallow and lower oceans, we can solve out for the reduced form model in
$M_{t}.$ Here, $M_{t}$ can be seen as following a Markov state-space process
with two hidden layers (latent variables)$.$ The system implies that the
reduced-form model for $M_{t}$ consists of an autoregressive model with
distributed lags of $E_{t},$ here an ARDL(4,4), which we log-linearize for
$m_{t}=\log M_{t}$ as
\begin{equation}
A\left(  L\right)  m_{t}=B\left(  L\right)  e_{t},
\end{equation}
where $L$ denotes the lag operator such that $Lm_{t}=m_{t-1}$ and $A\left(
\cdot\right)  ,$ $B\left(  \cdot\right)  $ are polynomials of degree 4. Now
atmospheric temperatures, $H_{t},$ themselves relate dynamically to
atmospheric gas concentrations, ocean temperatures, $H_{t}^{\left(
\ell\right)  }$ as well as extraneous radiative forcing, $F_{t}^{0}$, which in
log form can be written as
\[
h_{t+1}=\left(  1-a_{0}\right)  h_{t}+a_{1}m_{t+1}+a_{2}h_{t}^{\left(
\ell\right)  }+f_{t+1}^{0},
\]
where $\left[  1-\left(  1-a_{3}\right)  L\right]  h_{t}^{\left(  \ell\right)
}=a_{3}h_{t-1}$ specifies the temperature increase in the lower oceans. Then
the equation for temperature dynamics becomes
\begin{equation}
\left[  1-\left(  1-a_{3}\right)  L\right]  \left(  h_{t+1}-a_{1}%
m_{t+1}-f_{t+1}^{0}\right)  =\left\{  \left(  1-a_{0}\right)  \left[
1-\left(  1-a_{3}\right)  L\right]  +a_{2}a_{3}L\right\}  h_{t}.
\label{eq: temp dyn}%
\end{equation}
Finally, we consider the impact of economic growth: the fraction of GDP not
spent on abatement is consumed, $c_{t},$ or invested, $i_{t},$ along the
budget constraint:%
\begin{equation}
y_{t}-\omega_{t}-\xi h_{t}=\tilde{c}_{t}, \label{eq: y}%
\end{equation}
where $\tilde{c}_{t}=c_{t}+i_{t},$ and the logarithm of the cost of abatement,
$\omega_{t},$ satisfies%
\begin{equation}
\omega_{t}=\psi_{t}+\theta\mu_{t}.
\end{equation}
with $\theta>1$ so the cost of abatement increases faster than abatement
itself. Parameter $\xi$ in (\ref{eq: y}) represents damage induced by warming:
it closes the feedback loop from climate to the economy shown in Figure 1 in
the paper. Considering $\psi_{t}$ constant over the historical sample (but not
over the future under tipping point scenarios), we identify $\omega_{t}$ to
$\theta\mu_{t}$.

\section{Data construction}

Online links to sources and interpolation details are provided in separate
code file (\emph{Data Preparation}).

\paragraph{Economic Data}

We obtained real world output through the Angus Maddison historical database
currently maintained at the University of Gr\"{o}ningen (years AD 1 to 2009)
and linearly interpolated its logarithm ($y_{t}$). For lack of long series on
capital investment, we used as a proxy for $\tilde{c}_{t},$ the logarithm of
real world consumption provided in the Barro and Urs\`{u}a
(2010\nocite{barro2010barro}) Macroeconomic database. In this database, not
all series start at the same origin, so we backcast them to the early
nineteenth century, assuming that per capita consumption growth in individual
countries were equal to their subcontinental regions (as defined by Barro and
Urs\`{u}a). This allows to smoothen the breaks in the world consumption
series. For data prior to the beginning of the Urs\`{u}a-Barro data, 1820, we
assumed that consumption grew as the same pace as GDP (as measured in the
Maddison database).\ This strong assumption appears reasonable for the
pre-industrial era (despite heterogeneity that is manifest in Northern Europe
in particular, see Allen, 2001\nocite{allen2001great}).

\paragraph{Climate Data}

Climate variables are also obtained from various sources, where our focus is
primarily on using long series. Global atmospheric concentrations of GHG are
drawn from the Coupled Model Intercomparison Project Phase 6 (CMIP6)\ database
by Meinshausen \emph{et al.} (2017\nocite{CMIP6}). They date back to AD year 1
for a series of gasses, including CO$_{2}$. The data on Gas emissions dating
back to 1750 were obtained from the Global Carbon Project at OurWorldinData.org.

Temperature anomalies were obtained from several sources for robustness checks
and we produce two distinct long series. The first draws on 1500 Year Northern
Hemisphere Temperature Reconstructions by Hegerl \emph{et al.}
(2007\nocite{Hegerl07}), and we used the Northern Hemisphere Temperature
reconstruction calibrated to 30-90N land only which starts in AD 558 and stops
in 1960. \ We extend this series using anomalies from the raw land database
maintained by Berkeley Earth and which starts in 1750. To match both series we
regress the Hegerl \emph{et al.} series on lags of the Berkeley data and use
the resulting forecasts post 1960 as `reconstructed series'.\ We use this as
our main series below and denote it as TA, but we also check that our results
are robust to using the reconstruction by Mann \emph{et al.}
(2008\nocite{mann2008proxy}) labeled as \textquotedblleft2,000 Year
Hemispheric and Global Surface Temperature Reconstructions: Global: Land and
Ocean: Error-In-Variables Method\textquotedblright\ and denoted by TAman.

Finally, data on (non-)anthropogenic Radiative Forcing (RF) comes from two
sources. The series in Hegerl \emph{et al.} (2007\nocite{Hegerl07}) start in
year AD\ 1000 and covers volcanic as well as GHG origins.\ We complement them
with data from Crowley \& Unterman (2013\nocite{crowley2013technical}) who
provide detailed reconstruction of RF of volcanic origin (series AOD in
particular, with sign opposite that of Hegerl \emph{et al.}) available from AD 800.

\section{Empirical model}

Likelihood ratio tests for the rank of cointegration indicate the presence of
$r=2$ cointegration relations (with a Trace statistic of 31.69, corresponding
to an asymptotic $p$-value of 0.088), together with two \textsf{I}$\left(
1\right)  $ stochastic trends. The estimated cointegrating vectors are
obtained as (standard errors in parentheses when available, their absence
means that we have restricted the parameter for identification of the vectors):%

\begin{align}
\beta^{\prime}  &  =\left[  \left.
\begin{array}
[c]{cccc}%
y_{t} & c_{t} & m_{t} & h_{t}\\
\underset{\left(  -\right)  }{1} & \underset{\left(  .175\right)  }{-1.54} &
\underset{\left(  -\right)  }{0} & \underset{\left(  .399\right)  }{-.81}\\
\underset{\left(  -\right)  }{0} & \underset{\left(  -\right)  }{0} &
\underset{\left(  2.01\right)  }{-4.12} & \underset{\left(  -\right)  }{1}%
\end{array}
\right\vert
\begin{array}
[c]{c}%
1_{\left\{  t\geq1800\right\}  }t\\
\underset{\left(  .0053\right)  }{-.036}\\
\underset{\left(  -\right)  }{0}%
\end{array}
\right]  ,\\
\alpha^{\prime}  &  =\left[
\begin{array}
[c]{cccc}%
\Delta y_{t} & \Delta c_{t} & \Delta m_{t} & \Delta h_{t}\\
\underset{\left(  .00015\right)  }{-.0013} & \underset{\left(  .0004\right)
}{-.0014} & \underset{\left(  -\right)  }{0} & \underset{\left(  -\right)
}{0}\\
\underset{\left(  -\right)  }{0} & \underset{\left(  -\right)  }{0} &
\underset{\left(  5.78\mathsf{e}^{-5}\right)  }{1.33\mathsf{e}^{-5}} &
\underset{\left(  .0046\right)  }{-.024}%
\end{array}
\right]  .
\end{align}

The Likelihood Ratio test of overidentifying restrictions has a realization of
5.89, which corresponds to a $p$-value of 0.44 based on a $\chi_{6}^{2}.$ The
restrictions are therefore not rejected. We retain in the second adjustment
vector $\alpha$ the presence of $m_{t}$ though it is insignificant in the
whole sample. This ensures the system defines a proper statistical correction
towards equilibrium although the adjusting variable is temperature, not carbon
dioxide concentration.

\begin{figure}[ptb]
\centering\scalebox{.7}{\includegraphics{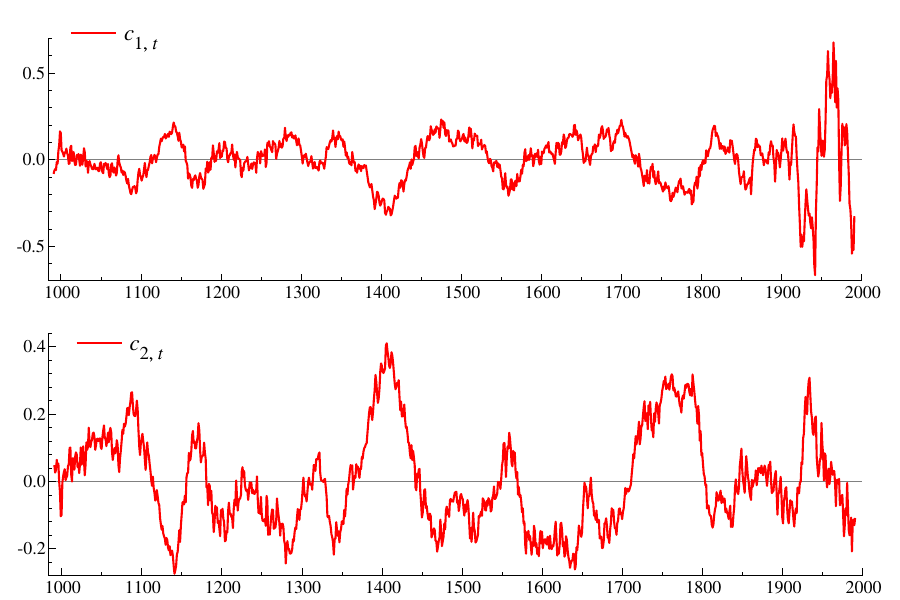}} \caption{The two
cointegrating vectors (corrected for short-run dynamics and deterministic
terms). $c_{1,t}$ refers to the interaction between the economy and
temperature. The second vector $c_{2,t}$ relates to the interaction between
CO$_{2}$ concentrations, exogenous radiative forcings and temperature. }%
\label{fig:coint}%
\end{figure}

The long run semi-elasticity of temperature to CO$_{2}$ concentrations is
4.12,\textbf{ }which implies that an increase of 100\% of concentrations
results, ceteris paribus, in a 2.9 Celsius/Kelvin increase in temperature
anomalies. Figure \ref{fig:coint} shows that the cointegrating vectors are
stable over long periods. This is particularly true of the second vector, but
despite additional volatility over the industrial revolution (in particular
during the initial transition that is roughly captured by our broken linear
trend), the first equation appears stable overall. 

\begin{figure}[ptb]
\centering\scalebox{.7}{\includegraphics{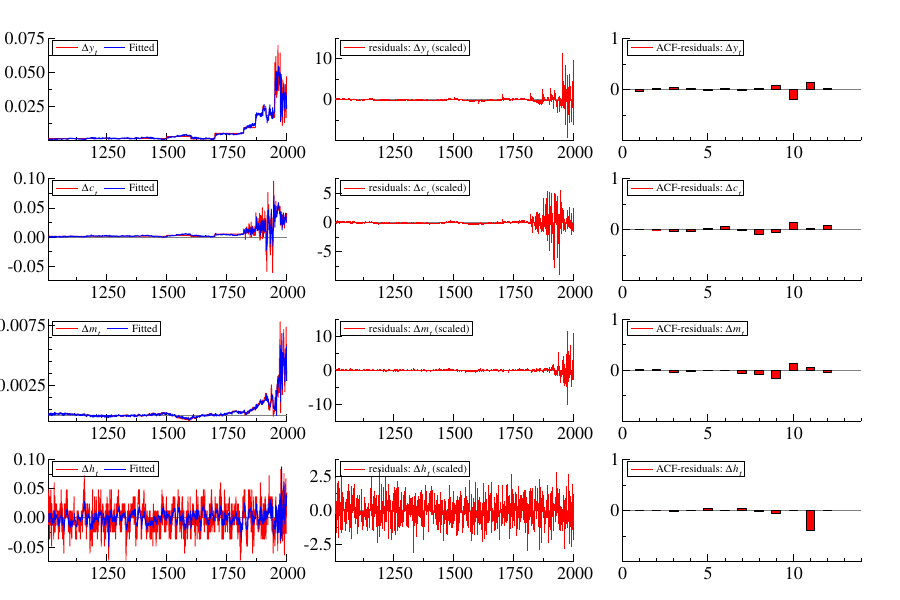}} \caption{Fit
of a VAR$\left(  8\right)  $ with two \textsf{I}$\left(  1\right)  $
stochastic trend.}%
\label{fig: modelcats}%
\end{figure}

Figure \ref{fig: modelcats} presents the fit of the estimated model in the
differences of the endogenous variables. The figure shows that the VAR(8)
misses important temperature autocorrelation at an 11 year-period, which might
correspond to the solar cycle. Also, the residuals exhibit a large shift in
volatility towards the end of the sample. The presence of an \textsf{I}$(2)$
stochastic trend is rejected when testing for it, so the mean shifts we
observe over the industrial revolution may likely be better model as a
location shift in the first difference (as in, e.g., Kim \emph{et al.},
2020\nocite{KOEP20}). The presence of severe heteroscedasticity. in the end of
the sample renders inference somewhat imprecise but we do not model this
change of volatility to avoid affecting the distribution of test statistics.

The corresponding $C$ matrix of long-run impact is estimated as
\[
\hat{C}=%
\begin{array}
[c]{cc}
& \mathbf{\color{blue}}\left[  y_{t}\quad\quad\quad c_{t}\quad\quad\quad
m_{t}\quad\quad\quad h_{t}\right]  \mathbf{\color{black}}\\
\mathbf{\color{blue}}\left[
\begin{array}
[c]{c}%
y\underset{}{_{\underset{}{t}}}\\
c_{\substack{\underset{}{t}\\}}\\
m_{\substack{\underset{}{t}\\}}\\
h_{\substack{\underset{}{t}\\}}
\end{array}
\right]  \mathbf{\color{black}} & \left[
\begin{tabular}
[c]{cccc}%
$\underset{\left[  3.31\right]  }{\mathbf{1.94}}$ & $\underset{\left[
-5.48\right]  }{-\mathbf{2.15}}$ & $\underset{\left[  -3.15\right]
}{-\mathbf{30.7}}$ & $\underset{\left[  -.243\right]  }{-.033}$\\
$\underset{\left[  1.4\right]  }{\mathbf{.}54}$ & $\underset{\left[
-3.16\right]  }{\mathbf{-.82}}$ & $\underset{\left[  -5.02\right]
}{-\mathbf{32.2}}$ & $\underset{\left[  -.228\right]  }{-.02}$\\
$\underset{\left[  3.69\right]  }{\mathbf{.33}}$ & $\underset{\left[
-4.45\right]  }{-\mathbf{.27}}$ & $\underset{\left[  3.83\right]
}{5\mathbf{.69}}$ & $\underset{\left[  -.023\right]  }{-.001}$\\
$\underset{\left[  \mathbf{\color{magenta}}3.69\right]  }%
{\mathbf{\color{magenta}1.36}}$ & $\underset{\left[  \mathbf{\color{magenta}}%
-4.45\right]  }{\mathbf{\color{magenta}-1.1}}$ & $\underset{\left[
\mathbf{\color{magenta}}3.83\right]  }{\mathbf{\color{magenta}23.4}}$ &
$\underset{\left[  -.023\right]  }{-0.002}$%
\end{tabular}
\ \ \right]  ,
\end{array}
\]
where figures in square brackets are t-statistics. The parameters and t-stats
magenta correspond to the test of long-term temperature controllability using
the controls given by their respective columns.

\paragraph{Comparison with the literature.}

Nordhaus \& Moffat (2017\nocite{NorMof17}) report projections that a $3%
{{}^\circ}%
$C increase in global temperature would reduce world GDP by $2.04\%,$ while a
$6%
{{}^\circ}%
$C increase would result in a 8.06\%\ damage. These figures imply that
parameter $\xi$ in equation equation (2c) would range from $.68$ to $1.34.$
Our corresponding estimate in the first cointegration relation is hence in the
lower range, at $\hat{\xi}=.81$, with a standard error of$\ 0.40.$

Assessing the role played by atmospheric carbon dioxide, Hassler, Krusell \&
Olovsson (2018\nocite{HKO18}) evaluate the joint effects of climate and the
economy. Focusing on climate sensitivity to carbon, and using the famous
Arrhenius equation, their result implies that the change in temperatures that
can be inputed to atmospheric carbon over the industrial revolution,
$h_{2000}-h_{1800},$ ranges between $h_{2000}-h_{1800}\in\left[
0.7,2.1\right]  .$ Our estimated long range elasticity $\partial
h_{t}/\partial m_{t}=4.12$ in the second cointegration relation, together with
a change of $m_{2000}-m_{1800}=.33,$ give an estimated value of temperature
change that can be inputed to CO$_{2}$ concentrations\ at $h_{2000}%
-h_{1800}=1.35$ with a standard error of .66, i.e., in the middle of the range
found in the literature.

Combining this effect with the interaction between temperatures and economic
activity, Hassler \emph{et al.}, (2018) calibrate $\gamma,$ the percentage
loss in world GDP due to a one-unit GtC (Giga tonnes of Carbon) increase in
$M_{t}$ as $\gamma\in\left[  .27,10.4\right]  .$ Our estimated model finds a
long run $\hat{\gamma}=4.25,$ using with a standard error of $0.375$, again in
the middle of the range of possible values entertained by these authors. To
obtain this value, we refer to Hassler, Krusell \& Olovsson
(2018\nocite{HKO18}). Under their data, the total amount of CO$_{2}$ in the
atmosphere in 2020 is 800\ GtC so a 1\% increase amounts to 8GtC. Our
estimated matrix $\hat{C}$ gives a long run elasticity of $y_{t}$ to $m_{t}$
of $-34$ so, for a 1 GtC increase, i.e., 0.125\% of 2020, we obtain an
estimated $\hat{\gamma}=4.25$ with corresponding standard deviation.

\section{Invariance}

Invariance refers to the property that the model is not affected by the
introduction of the new policy rule. It relates to the concept of
super-exogeneity by Engle, \emph{et al., }(1983\nocite{engle1983exogeneity}).
The difference here is that we do not rely only on invariance of the
conditional model for $h_{t}$ (and possibly $m_{t}$) given $\left(
y_{t},c_{t}\right)  $ to changes in the determination of the economic
variables, but on the stability of the system. For this reason, we ascertain
its presence through two approaches:

\begin{enumerate}
\item In our empirical application, we use very long spans of data that allow
us to capture relations that are stable pre- and post-industrial revolution
(as seen in Figure \ref{fig:coint}). Hence our maintained assumption is that
cointegrating vectors and their impacts $\left(  \alpha,\beta\right)  $ remain
stable. The presence of a constant cointegrating vectors over the subsamples
corroborates this assumption (though we know that reconstructed dataset may
generate misleading results, see Pretis \& Hendry, 2013\nocite{esd-4-375-2013}%
). In order to give support to this assumption, we checked the stability of
the estimated cointegrating vectors over the sample period using the method of
Hansen \& Johansen (1999\nocite{HS1999}); Figure \ref{fig: beta_const} below
displays a path of the stability test over the sample, which lies below the
95\% quantile of the limit distribution (adopting a restricted-linear trend
specification, see the subsection below for further details of the test).
Although we carefully introduced stochastic trends in the model (such as that
in $\sigma_{t}$) to account for economic development and the time-varying
energy mix, it may be difficult to justify in an empirical context that all
the parameters are constant over our very long sample. Yet, the evidence of
stability recorded in Figure \ref{fig: beta_const} is nonetheless encouraging
for us in employing this cointegrated model for counterfactual
policy-simulation purposes.

\item We also use techniques derived for tests of super-exogeneity, as
proposed by Hendry and Santos (2010\nocite{HendSant10}) and discussed in
Castle,\ Hendry \& Martinez (2017\nocite{Castle17}). These rely on modeling
the marginal distributions of the endogenous variables, and detecting
corresponding outliers in the errors. These outliers constitute possible
shifts or breaks that are then introduced as regressors in the conditional
equations. If the outliers enter significantly, this implies that the
conditional distributions are not immune to changes in the marginal equations,
thus generating a failure of super-exogeneity. Hendry and Santos perform an
automated test via impulse saturation (one possible outlier per observation)
and selection (Santos, Hendry \& Johansen, 2008\nocite{santos2008automatic}).
Owing to residual heteroscedasticity in our model over the industrial era,
outlier detection is not feasible but we replace the corresponding dummy
variables with exogenous radiative forcings (solar and volcanic, as obtained
from Hegerl \emph{et al.}, 2007). Using automated model selection (see Castle
\emph{et al.}, 2012 for a discussion of the Autometrics
algorithm\nocite{castle2012model}) for both the marginal and conditional
models (in first differences, and with a 1\% size for search), we find that
several volcanic RF variables enter the marginals, but none enters the
conditional model, thus indicating a rejection of the hypothesis of no invariance.
\end{enumerate}

\paragraph{Stability of cointegrating parameters}

We make a brief review of the test statistic for stability of cointegrating
parameters discussed above. We employ the Nyblom-type test (Nyblom,
1989\nocite{Nyblom1989}) developed by Hansen \& Johansen (1999\nocite{HS1999})
and adopt the formulation of Br\H{u}ggeman, Donati \& Warne
(2003\nocite{BDW2003}) to improve its numerical accuracy. Letting $R_{0t}$ and
$R_{1t}\ $denote a set of residuals from auxiliary regressions in the
concentrated likelihood function (see Johansen,\ 1996, chap. 6), a sequence of
sample product moments is obtained, for $t=1,...T,$ as
\[
\left(
\begin{array}
[c]{cc}%
S_{00}^{(t)} & S_{01}^{(t)}\\
S_{10}^{(t)} & S_{11}^{(t)}%
\end{array}
\right)  =\frac{1}{t}\sum\limits_{i=1}^{t}\left(
\begin{array}
[c]{cc}%
R_{0i}R_{0i}^{\prime} & R_{0i}R_{1i}^{\prime}\\
R_{1i}R_{0i}^{\prime} & R_{1i}R_{1i}^{\prime}%
\end{array}
\right)  .
\]
We then compute a sequence of test statistics, for $t=1,...,T$ as
\[
L_{T}^{\left(  t\right)  }=\left(  \frac{t}{T}\right)  ^{2}tr\left\{  \left(
V^{(T)}\right)  ^{-1}S^{\left(  t\right)  \prime}\left(  M^{(T)}\right)
^{-1}S^{\left(  t\right)  }\right\}  ,
\]
where $V^{(T)}=\widehat{\alpha}^{\prime}\left(  \widehat{\Omega}\right)
^{-1}\widehat{\alpha},$ $S^{\left(  t\right)  }=N^{\prime}\left(  S_{01}%
^{(t)}-\widehat{\alpha}\widehat{\beta}^{\ast\prime}S_{11}^{(t)}\right)
^{\prime}\left(  \widehat{\Omega}\right)  ^{-1}\widehat{\alpha}$ and
$M^{(T)}=T^{-1}N^{\prime}S_{11}^{(T)}N$ for some suitable normalization matrix
$N$. Comparing $L_{T}^{\left(  t\right)  }\ $with a reference based on the
limit distribution of its supremum allows us to check parameter stability.
Note that $L_{T}^{\left(  t\right)  }$ simply uses a class of full sample
estimates, $\widehat{\alpha}$,$\ \widehat{\beta}^{\ast}\ $and$\ \widehat
{\Omega}$ and involves no recursive estimation, so that the sequence of
$L_{T}^{\left(  t\right)  }\ $can be calculated over $t=1,...,T$ as shown in
Figure \ref{fig: beta_const}. The 95\% quantile of the limit distribution
obtained from the \textit{CATS} manual is used as a reference in the figure.

\begin{figure}[t]
\centering\scalebox{.7}{\includegraphics{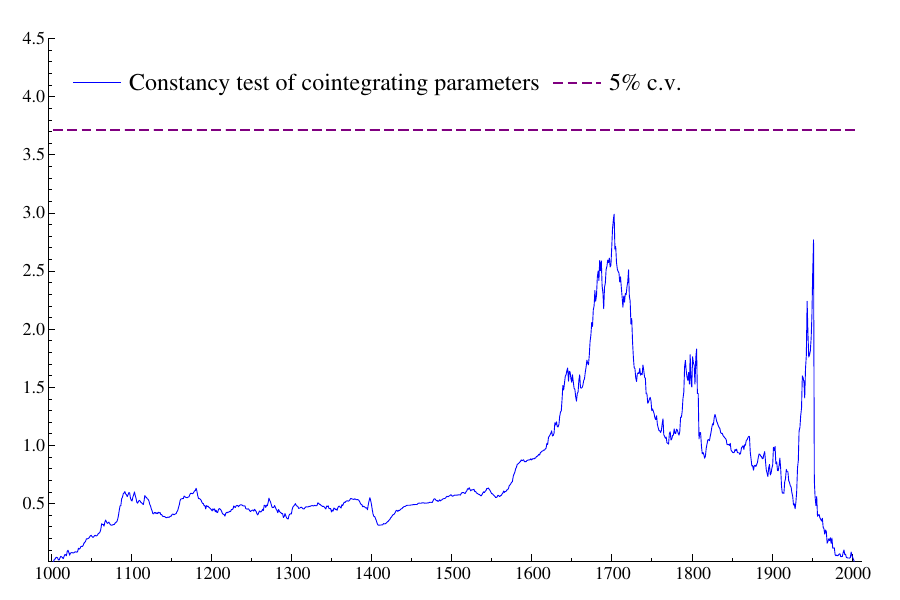}}\caption{Testing
for constancy of cointegrating parameters: the figure reports the sequence of
test statistics for the null of cointegration parameter constancy.}%
\label{fig: beta_const}%
\end{figure}

\section{Comparison with Rambachan and Shephard (2021)}

In the Rambachan and Shephard (2021\nocite{rambashep21}) framework, at each
period $t\geq1,$ the unobserved unit $X_{t}$ receives a random assignment
$W_{t}$ and we observe an outcome $X_{t}^{new}\left(  W_{t}\right)  .$ The
\textquotedblleft potential outcome process\textquotedblright\ at time $t$,
for any deterministic sequence $\left\{  w_{s}\right\}  ,$ is $X_{t}%
^{new}\left(  \left\{  w_{s}\right\}  _{s\geq1}\right)  .$ Under the
assumption of \emph{Non-anticipating Potential Outcomes}, for each $t\geq1$
and all deterministic sequences $\left\{  w_{t}\right\}  _{t\geq1},$ $\left\{
w_{t}^{\prime}\right\}  _{t\geq1}$%
\[
X_{t}^{new}\left(  w_{1:t},\left\{  w_{s}\right\}  _{s\geq t+1}\right)
\overset{a.s.}{=}X_{t}^{new}\left(  w_{1:t},\left\{  w_{s}^{\prime}\right\}
_{s\geq t+1}\right)  .
\]
They make the link with the macroeconomic literature on impulse response
functions (IRF), defined in the context of Structural VARs as (Sims \emph{et
al.}, 1982\nocite{sims1982policy})
\[
\mathsf{IRF}_{k,t,h}\left(  w_{k},w_{k}^{\prime}\right)  \equiv\mathsf{E}%
\left[  \left.  Y_{t+h}\left(  W_{k,t}\right)  \right\vert W_{k,t}%
=w_{k}\right]  -\mathsf{E}\left[  \left.  Y_{t+h}\left(  W_{k,t}\right)
\right\vert W_{k,t}=w_{k}^{\prime}\right]  .
\]
Rambachan and Shephard (2021)\ show that the IRF can be be given a causal
meaning, coinciding with the Average Treatment Effect $\mathsf{E}\left[
Y_{t+h}\left(  w_{k}\right)  -Y_{t+h}\left(  w_{k}^{\prime}\right)  \right]  $
under some orthogonality conditions that are satisfied when the assignment
constitutes a \textquotedblleft shock\textquotedblright, defined as in their
Theorem 2 when $W_{k,t}\perp\left(  W_{1:t-1},W_{k^{\prime},t},W_{t+1:t+h}%
,\left\{  X_{t+h}^{new}\left(  w_{1:t+h}\right)  \right\}  \right)  .$ From
equation (8) in the main text, we see that the assignment $W_{t}$
corresponding to the control policy is zero when the data generating process
for $X_{t}^{new}$ coincides with that of $X_{t}.$ We can there define the
assignment as
\begin{equation}
W_{t+1}=\overline{a}\nu\left(  X_{t}^{new}\right)  . \label{Assignment}%
\end{equation}
In our context where the policy is implemented at every period, Rambachan and
Shephard (2021\nocite{rambashep21}) define the \emph{impulse causal effect at
horizon }$h\geq1$\emph{ }as the difference between $X_{t+h}^{new}$ and the
\emph{counterfactual} $X_{t+h}^{\ast new}$ that obtains with the only change
that the policy is not implemented at time $t$ (so $W_{t}=\nu\left(
X_{t-1}^{\ast}\right)  =0$ under the counterfactual and this is the only
difference in assignments). In their words, the impulse causal effect measures
the ceteris paribus causal effect of intervening to switch the time-$t$
assignment from 0 to $W_{t}$ on the $h$-period ahead outcomes holding all else
fixed along the assignment process. Since $X_{t}$ is non-stationary, the
impulse causal effect and its unconditional expectation, the Average Treatment
Effect, vary with time. Yet, we notice that the policy intervention, $W_{t+1}$
in (\ref{Assignment}) does not constitute a contemporaneous \textquotedblleft
shock\textquotedblright\ in the Ramey (2016\nocite{Ramey16}) or Rambachan and
Shephard (2021, Theorem 2\nocite{rambashep21}) sense, since $W_{t+1}$ is not
unanticipated from, or uncorrelated with, lagged endogenous variables, in fact
it is possibly persistent (though stationary under the assumption of
controllability, Condition \textsf{C}). In practice, JJ, Theorem 6, show there
exists a linear policy rule that ensures $W_{t+1}$ can be expressed as a
function of the lagged shocks to the unperturbed system and can be made $iid$.
In the context of the VAR$\left(  1\right)  ,$ following on the implementation
of the policy, the data generating process writes%
\begin{equation}
X_{t+1}^{new}=-\left[  \alpha\mu+\left(  I_{p}+\alpha\beta^{\prime}\right)
\overline{a}\kappa_{0}\right]  +\left(  I_{p}+\alpha\beta^{\prime}\right)
\left(  I_{p}+\overline{a}\kappa^{\prime}\right)  X_{t}^{new}+\varepsilon
_{t+1} \label{eq: VARnew}%
\end{equation}
so $\kappa^{\prime}X_{t+1}^{new}=-\kappa^{\prime}\alpha\mu+\kappa^{\prime
}\left(  I_{p}+\alpha\beta^{\prime}\right)  \left(  \left(  I_{p}+\overline
{a}\kappa^{\prime}\right)  X_{t}^{new}-\overline{a}\kappa_{0}\right)
+\kappa^{\prime}\varepsilon_{t+1}.$ Under policy assumptions $\kappa^{\prime
}\alpha=0$ and $I_{m}+\kappa^{\prime}\overline{a}=0.$ The previous expression
then simplifies as%
\[
\kappa^{\prime}X_{t+1}^{new}=\kappa^{\prime}\left(  \left(  I_{p}+\overline
{a}\kappa^{\prime}\right)  X_{t}^{new}-\overline{a}\kappa_{0}\right)
+\kappa^{\prime}\varepsilon_{t+1}=\kappa_{0}+\kappa^{\prime}\varepsilon_{t+1}%
\]
i.e., $W_{t+1}=\overline{a}\nu\left(  X_{t}^{new}\right)  =\overline{a}%
\kappa^{\prime}\varepsilon_{t}$. Hence, the data generating process under the
new policy -- the process for the potential outcome -- becomes%
\begin{align}
\Delta X_{t+1}^{new}  &  =\alpha\left(  \beta^{\prime}X_{t}^{new}-\mu\right)
+\left(  I_{p}+\alpha\beta^{\prime}\right)  W_{t+1}+\varepsilon_{t+1}%
\nonumber\\
&  =\alpha\left(  \beta^{\prime}X_{t}^{new}-\mu\right)  +\left(  I_{p}%
+\alpha\beta^{\prime}\right)  \overline{a}\kappa^{\prime}\varepsilon
_{t}+\varepsilon_{t+1}, \label{eq: VARMA}%
\end{align}
a VARMA$\left(  1,1\right)  .$ Using the results in Theorem 6 of JJ, we can
show the same result for a VAR$\left(  p\right)  $ that becomes a
VARMA$\left(  p,1\right)  $ under the policy. Hence, since $\varepsilon
_{t}\perp\varepsilon_{t+1}$ an alternative SVAR representation is feasible
\begin{equation}
X_{t+1}^{new}=-\alpha\mu+\left(  I_{p}+\alpha\beta^{\prime}\right)
X_{t}^{new}+B\underline{\varepsilon}_{t+1} \label{eq: SVAR}%
\end{equation}
where $\underline{\varepsilon}_{t+1}=\left(  \varepsilon_{t+1}^{\prime
},\varepsilon_{t}^{\prime}\right)  ^{\prime}$ and $B$ conforms with
(\ref{eq: VARMA}). In equation (\ref{eq: SVAR}), $\underline{\varepsilon
}_{t+1}$ contains an excess shock that is recoverable from past observations
since by construction $\kappa^{\prime}X_{t+1}^{new}$ induces an additional
cointegration relation so the eigenvalues of $\left(  I_{p}+\alpha
\beta^{\prime}\right)  \overline{a}\kappa^{\prime}$ in (\ref{eq: VARnew}) --
and hence also in (\ref{eq: VARMA}) -- have modulus strictly smaller than unity.

\section{Additional counterfactual evidence}

\begin{figure}[ptb]
\centering\scalebox{.7}{\includegraphics{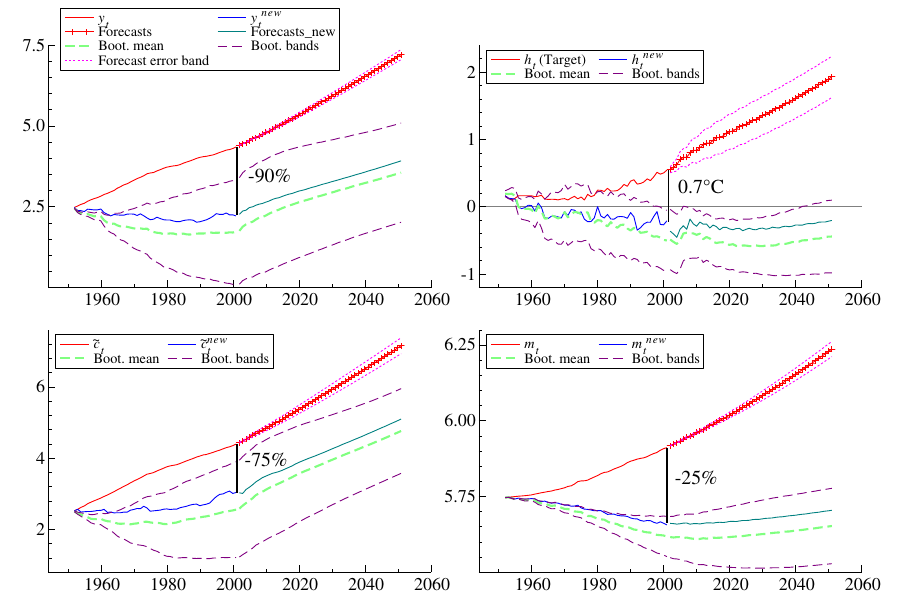}}\caption{Counterfactual
cost of a policy starting in 1950\ aiming at controlling global temperatures
and rendering them stationary around a long run mean equal to their 1900 level
(as known now).}%
\label{fig: Policy1950}%
\end{figure}

Figure \ref{fig: Policy1950} presents\ a complementary counterfactual analysis
where the objective in terms of temperature control remains the same, but
Policy 1 only starts in 1950. The Figure shows the cost of inaction in the
face of climate change is very large as the cost is now on average 90\% \ and
75\%\ of 2008 global GDP and consumption respectively (as opposed to 75\% and
45\% if the policy starts in 1990), as CO$_{2}$ concentrations need to be
reduced by 25\% of their 2008 levels (20\% in the baseline scenario).

\renewcommand{\baselinestretch}{1.1} \selectfont
\bibliographystyle{chicago}
\bibliography{Climatebiblio}